\documentclass[11pt]{article}

\usepackage{graphicx}
\graphicspath{{FIGS/}}

\newcommand{\comment}[1]{}

\usepackage{xcolor}

\usepackage{fancyvrb}

\usepackage{xspace}
\newcommand{\ie}{i.e.\@\xspace}
\newcommand{\eg}{e.g.\@\xspace}
\newcommand{\vs}{\textit{vs.}\@\xspace}

\pdfoutput=1

\usepackage[margin=0.9in]{geometry}
\usepackage{graphicx, amsmath, url,courier,array,csquotes}

\usepackage{authblk}

\usepackage{hyperref}

\newcommand{\e}{\mathrm{e}}

\usepackage{booktabs}
\usepackage{ragged2e}

\usepackage{tabularx}

\newcommand{\citep}{\cite}

\begin{document}








\title{GeneSurrounder: network-based identification of disease genes in 
expression data}
          

\renewcommand\Authands{ and }
\renewcommand\Authfont{\large}
\renewcommand\Affilfont{\small\itshape}

\author[1]{Sahil D. Shah}
\author[1,2]{Rosemary Braun\thanks{Corresponding author: \texttt{rbraun@northwestern.edu}}}

\affil[1]{Engineering Sciences and Applied Mathematics}
\affil[2]{Biostatistics, Feinberg School of Medicine and Northwestern Institute on Complex Systems}

\affil[ ]{Northwestern University}

\date{\today}

\maketitle

\abstract{
\noindent


\textbf{Background} 
A key challenge of identifying disease–associated genes is analyzing
transcriptomic data in the context of regulatory networks that control cellular
processes in order to capture multi-gene interactions and yield mechanistically
interpretable results. One existing category of analysis techniques  identifies
groups of related genes using interaction networks, but these gene sets often
comprise tens or hundreds of genes, making experimental follow-up challenging. 
A more recent category of methods identifies  precise gene targets while
incorporating systems-level information, but these techniques do not determine
whether a gene is a driving source of changes in its network, an important
characteristic when looking for potential drug targets.

\textbf{Results} 
We introduce GeneSurrounder, an analysis method that
integrates expression data and network information in a novel procedure to
detect genes that are sources of dysregulation on the network. The key idea of
our method is to score genes based on the evidence that they influence the
dysregulation of their neighbors on the network in a manner that impacts cell
function. Applying GeneSurrounder to real expression data, we show that our
method is able to identify biologically relevant genes, integrate pathway and
expression data, and yield more reproducible results across multiple studies of
the same phenotype than competing methods.

\textbf{Conclusions} 
Together these findings suggest that GeneSurrounder
provides a new avenue for  identifying individual genes that can be targeted
therapeutically. The key innovation of GeneSurrounder is the combination of
pathway network information with gene expression data to determine the degree to
which a gene is a source of dysregulation on the network. By prioritizing genes
in this way, our method provides insights into disease mechanisms and suggests
diagnostic and therapeutic targets. Our method can be used to help biologists
select among tens or hundreds of genes for further validation. The
implementation in R is available at github.com/sahildshah1/gene-surrounder

}

\maketitle


\section{Background} 



The advent of high--throughput transcription profiling technologies has enabled
identification of genes and pathways associated with disease, providing new
avenues for precision medicine. A key challenge is to analyze this data in the
context of the regulatory networks that control cellular processes,
while still obtaining insights that can be used to design new diagnostic and
therapeutic interventions.  It is thus necessary to develop methods that analyze
omic data in the context of the full network of interactions, while still
providing specific, targetable gene-level findings.


The most common method for detecting gene-association is via differential
expression analysis, in which each gene is independently tested for significant
differences in mean expression between phenotypes~\citep{Smyth2005}. However,
while differential expression analysis can identify specific (and hence
targetable) disease-associated genes, it does not  take into consideration the
network of molecular interactions that  govern cellular function, limiting the
mechanistic insights that can be derived from the data.  As a result, this
analysis can  miss crucial multi-gene interactions that underlie
complex phenotypes. Since biological systems are complex and expression data
is typically noisy, the multi-gene mechanisms that underlie a disease may be
detectable across multiple studies, but the individual genes contributing to
those mechanisms may vary from one study to the next. As a result, 
differential expression analysis can exhibit poor agreement between different
studies of the same conditions~\citep{MANO06,Khatri2012,SHAH2014}.

Maps of experimentally derived molecular interaction networks contained in
pathway databases and the growth of analysis techniques that
infer context-specific interaction networks have enabled the development of methods
that integrate systems level information with expression data. 
KEGG~\citep{Kanehisa2008}, for example, is a well-established pathway database that organizes
genes into hundreds of individual networks corresponding to biological processes. 
One use of interaction networks has been to identify \textit{groups} of related
genes underlying a biological mechanism.  By incorporating systems-level
information, these  pathway analysis techniques  can capture
multi-gene interactions, yielding mechanistically interpretable results that are
more reliable than single--gene analyses~\citep{MANO06,Khatri2012,SHAH2014}.
Pathway analysis techniques  can be broadly grouped into three categories:
`functional scoring methods', `topology methods', and `active modules tools.'
Functional scoring methods, such as GSEA~\citep{Subramanian2005}, identify
groups of genes that are enriched for association with the phenotype of
interest. Topology methods, such as SPIA~\citep{Tarca2009-wj} and
CePa~\citep{Gu2013-px,Gu2012-qo}, also identify groups of genes that are enriched
for association, but augment functional scoring methods with additional
information about the network of interactions between the genes. Active modules
tools, such as jActiveModules~\citep{Ideker2002}, HotNet~\citep{Vandin2011-zg},
and COSINE~\citep{Ma2011-ag}, attempt to find disease associated subnetworks
within pathways. These methods integrate systems-level information with
expression data to identify groups of related genes.


While pathway analysis techniques integrate systems--level information with omic
data to provide functional interpretations of the dataset, the ``significant
pathways'' identified by such analyses often comprise tens or hundreds of genes,
making experimental follow-up challenging. Additionally, boundaries between
pathways are often arbitrary, thus potentially neglecting key interactions. Moreover, many techniques rely on user--settable
parameters and ad-hoc heuristics that depend on network size, limiting their
interpretability and reliability~\citep{SHAH2014,jiang2014assessment}. Together,
these issues point to the need for analysis techniques that integrate network
and omics data to identify \textit{precise} gene targets for  follow-up studies.



Early efforts to identify precise gene targets while incorporating
systems-level information include
ENDEAVOUR~\citep{Aerts2006-in} and GeneWanderer~\citep{Kohler2008-nr}. ENDEAVOUR
takes in as input various data sources (such as literature abstracts and
protein-protein interactions) and prioritizes genes based on their similarity to
genes known to be involved in the disease. GeneWanderer uses protein-protein
interaction networks and identifies gene targets based on distance to known
disease genes on the network. However, these methods require knowledge of
mechanisms known to be associated with the disease.
Later analysis techniques -- such as a method that uses the Laplacian
kernel~\citep{Nitsch2009}, an extension of SPIA~\citep{Shafi2015-qu}, and
nDGE~\citep{Wu2013-tp} -- addressed this issue and do not require knowledge of
disease associated mechanisms to identify precise gene targets. The first method
uses a protein association network, recomputes distances using the Laplacian
kernel, and finds disease genes based on ``neighboring'' differential expression.
Since the distances are recomputed, the neighbors could include genes that are not neighbors on the original network. In other words, this method uses indirect
interactions instead of direct interactions, complicating the interpretation. 
In the extension of SPIA~\citep{Shafi2015-qu}, disease genes are found by
propagating changes in expression along the edges of the individual pathway so
that each gene is scored for disease-association according to its own change in
expression combined with the change in expression of its upstream neighbors.
Since each pathway is considered separately and the pathways have artificial
(sometimes overlapping) boundaries, an individual pathway could exclude genes
that are on a global network (\ie union of the individual pathways).
nDGE takes in as input expression profiles and for a putative disease gene class
conditionally identifies its co-regulated and actively co-regulated neighbors.
While powerful, each of these is limited in its treatment of the networks.
These methods either do not consider direct interactions between genes
on a global network (\citep{Nitsch2009} uses indirect interactions based on the
Laplacian kernel and~\citep{Shafi2015-qu} considers each KEGG pathway
separately) or infer interactions based on correlations (\eg, \citep{Wu2013-tp}).
Thus due to the limitations of the previously described techniques, an analysis
technique that takes into account direct interactions between genes globally
may prove useful in identifying targets and the effect they have on the network. 

Most recently, LEAN~\citep{Gwinner2016-kq} was developed to use direct
interactions on a global interaction network and find disease genes by scoring
the differential expression of ``local subnetworks.'' 
LEAN scores each gene for disease-association according to the enrichment of its
immediate neighbors. Thus, LEAN's algorithm restricts its focus to a local
subnetwork that only considers nearest neighbors. As a result, LEAN only
identifies genes based on the changes in expression of a given gene's local
subnetwork, but can not determine whether that gene is actually the source of
changes in its neighborhood or on the network; an important characteristic when
looking for potential targetable disease genes for use in precision medicine.


The goal of the present work is to combine  pathway network information with
gene expression data to determine the degree to which a gene is a source of
dysregulation on the network. 
We present a novel analysis technique, GeneSurrounder, that takes into account
the complex structure of interaction networks to identify specific disease-associated genes from
expression data. 
The key idea of our method is to score genes based on the evidence that they
influence the dysregulation of their neighbors on the network in a manner
that impacts cell function. In this way, the genes
returned by our method may be considered sources of ``disruption'' on the network and
therefore candidate targets for therapeutics. We thus seek to identify genes with two defining characteristics: $(i)$
they appear to influence other genes nearby in the network, as evidenced by
strongly correlated expression with nearby genes; and $(ii)$ their dysregulation
is associated with disease, as evidenced by a pattern of (progressively weaker) differential
expression centered about that gene. By finding these genes, our method 
identifies candidate genes that are ``disruptive'' to the mechanisms underlying
a given phenotype and does so without any reliance on user-set parameters or
arbitrary pathway boundaries.


In this manuscript, we describe the GeneSurrounder algorithm and  apply it to
data from three independent studies of ovarian cancer to demonstrate its use,
evaluate the reproducibility of its results, and demonstrate the methodological
and biological validity of our approach.  In order to evaluate the algorithm, we
evaluate its cross-study concordance, \ie, its consistency across different
data sets measuring the same phenotype.  We compare the cross-study concordance
of GeneSurrounder's results to that of standard differential expression
analysis, and find that genes identified as sources of pathway disruption by
GeneSurrounder are more consistently identified across the various studies than
are differentially expressed genes.  We also compare  our method to LEAN, and
show that genes identified by GeneSurrounder are 
more consistent across studies than both LEAN and differential expression analysis. 
We demonstrate that our method represents an integration of pathway and expression
data to yield results that are not solely driven by either alone 
and find that it identifies genes associated with ovarian cancer. Together,
these results suggest that GeneSurrounder reproducibly detects 
functionally-relevant genes by integrating  gene expression and network data. 
Our novel analysis approach complements existing gene-- and pathway--based
analysis strategies to identify specific genes that control disease--associated
pathways, providing a new strategy for identifying promising therapeutic
targets. 


\section{Methods} 


Our goal is to identify candidate
disease genes by analyzing gene expression data in the context
of interaction networks to discover genes that drive the behavior of pathways
associated with disease. We thus seek to identify genes with two defining
characteristics: $(i)$ they appear to influence other genes nearby in the
network, as evidenced by strongly-correlated expression with nearby genes; and
$(ii)$ their dysregulation is associated with disease, as evidenced by a pattern
of differential expression centered about that gene. Since the `extent'
of dysregulation of a given gene on a global gene network is not
known \textit{a priori}, we score the gene separately for every neighborhood
size on the
network (i.e. genes one `hop' away, genes up to two `hops' away, etc) and then return the results for the highest scoring neighborhood.
Genes with significantly high-scoring neighborhoods may then be prioritized for
follow-up experiments. 

\comment{ 
Our analysis technique identifies candidate disease genes by integrating
systems-level information with gene expression data. We developed our method to
find the genes with neighbors on the network that are differentially expressed
(with the magnitude of the differential expression decreasing with distance from
the putative disease gene) and have correlated expression with the putative
disease gene. Since the differential expression of the neighbors of a putative
disease gene does not depend on their association with that gene, our algorithm
}

\begin{figure}[!hb]
\centering
\includegraphics[width=\columnwidth]{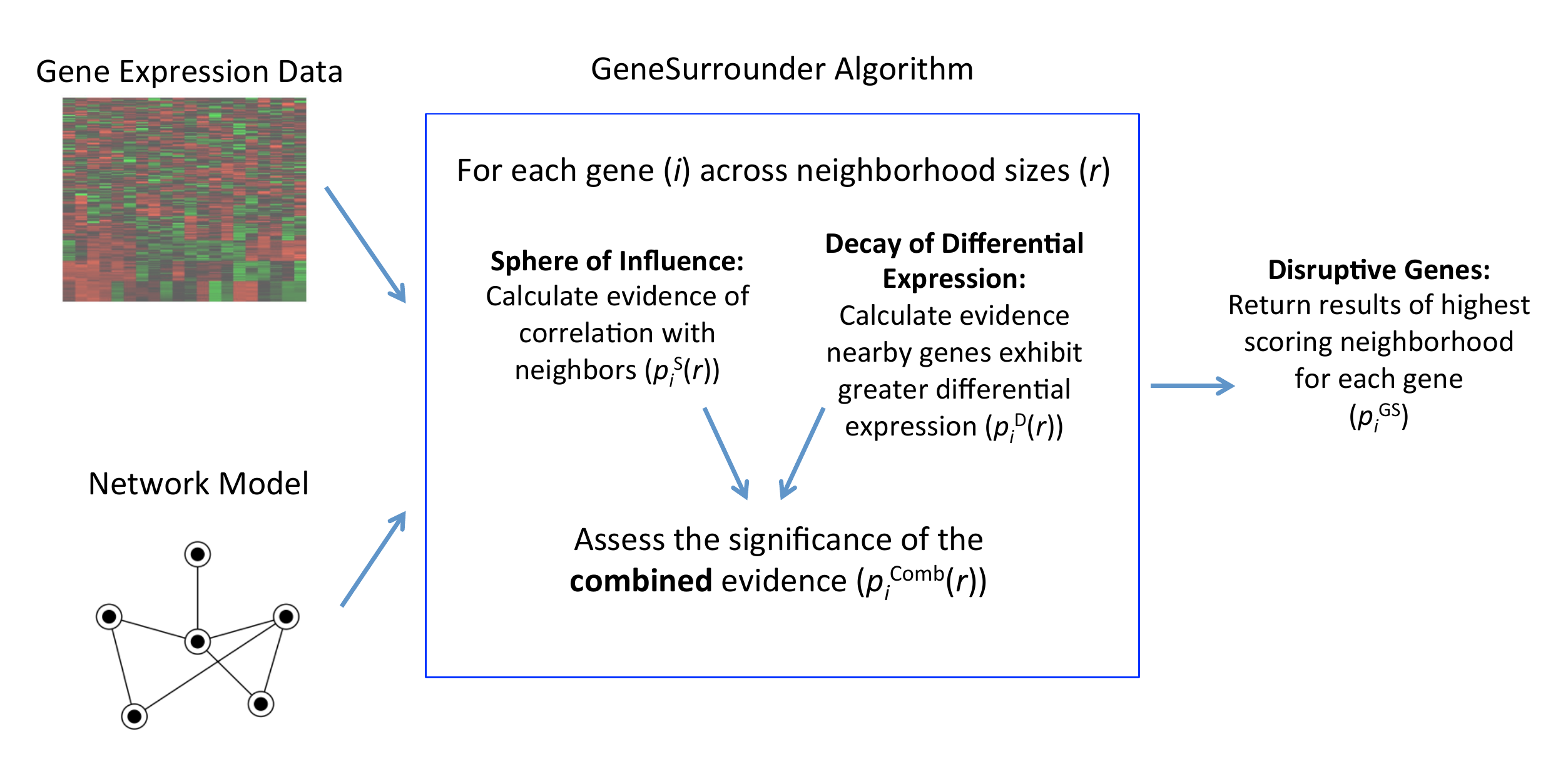}

\caption{\textbf{Overview of GeneSurrounder algorithm.} The algorithm incorporates 
systems--level information, in the form of a network model of cellular
interactions, with gene expression data to identify the genes that control
disease--associated mechanisms. The algorithm than identifies
``disruptive'' genes by assessing the significance of the combined evidence 
that (1) a gene has a influence on others in the network 
and (2) that its influence is driving disease.}

\label{fig:overview} 
\end{figure}

To this end, the GeneSurrounder method  consists of two tests that are run
independently of each other  (Figure~\ref{fig:overview}) and then combined, for
every neighborhood size on the network. To
determine if the putative disease gene is a ``disruptive'' candidate disease
gene meeting both criteria, the results for the  highest scoring
neighborhood are returned. 
%
%
To prioritize genes, our method is applied exhaustively to each assayed gene in
a transcriptomic data set, and the results from each gene's highest scoring
neighborhood are compared to rank the genes.

The algorithm takes as input gene expression data and a  network model of
cellular interactions derived from a pathway database.  In order to consider the
full scope of a gene's interactions and avoid artificially  imposed pathway
boundaries, we create a global KEGG network by merging the individual pathways
so that the links which are in at least one KEGG pathway will be part of the new
global network (\ie, the graph union of all pathways). 
We then consider the largest connected component of the
resulting network in our algorithm. Using this global network
and gene expression data, we compute evidence for each of the above criteria
as follows.

\begin{figure}[hbt]

\centering
\includegraphics[width=0.75\columnwidth]{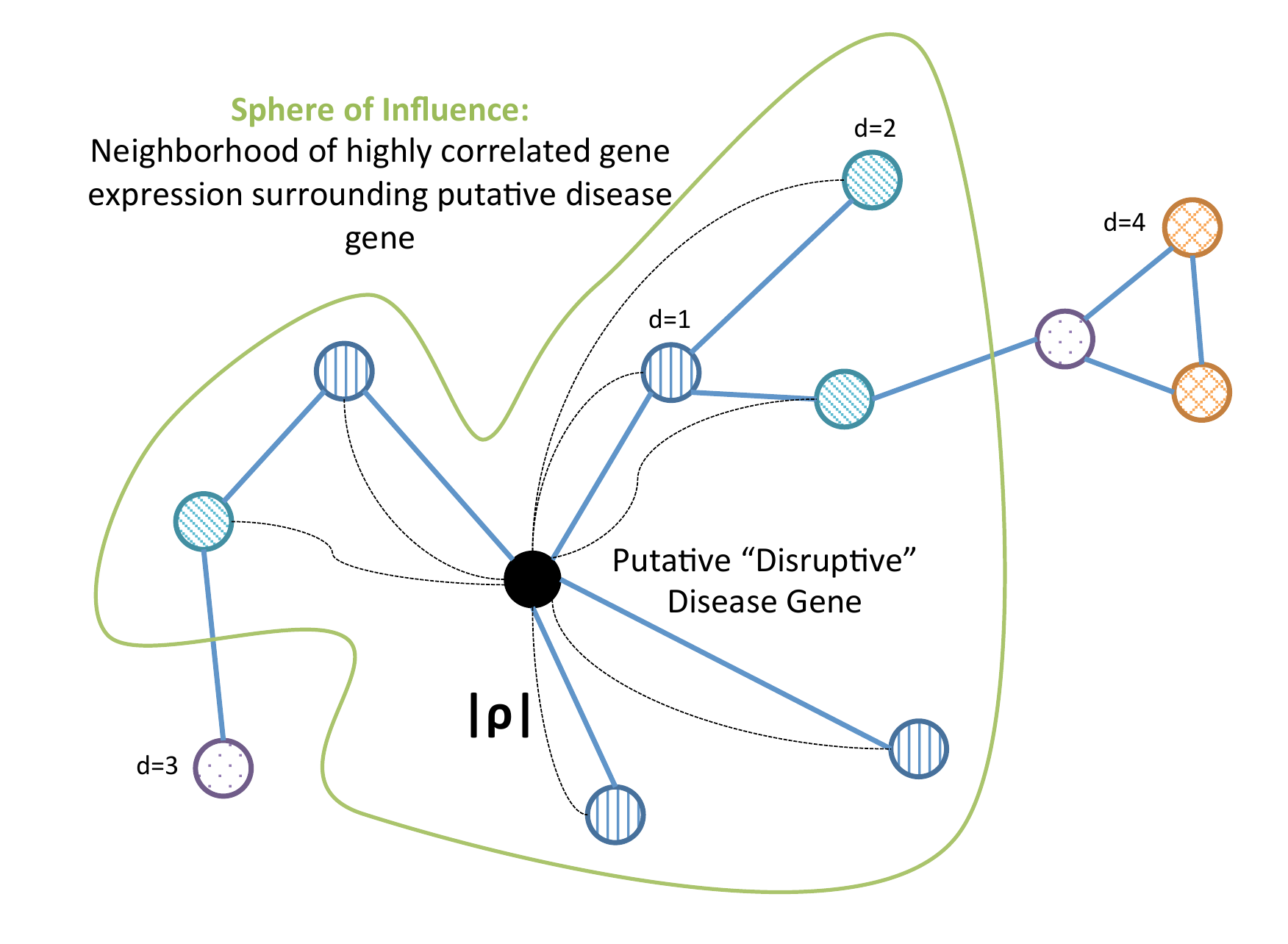}

\caption{\textbf{Procedure for Sphere of Influence.} The Sphere of Influence computation
tests if a putative driver gene is more correlated with its neighbors
than a random sample of genes.}

\label{fig:f1}
\end{figure}

\begin{figure}[hbt]
\centering
\includegraphics[width=0.75\columnwidth]{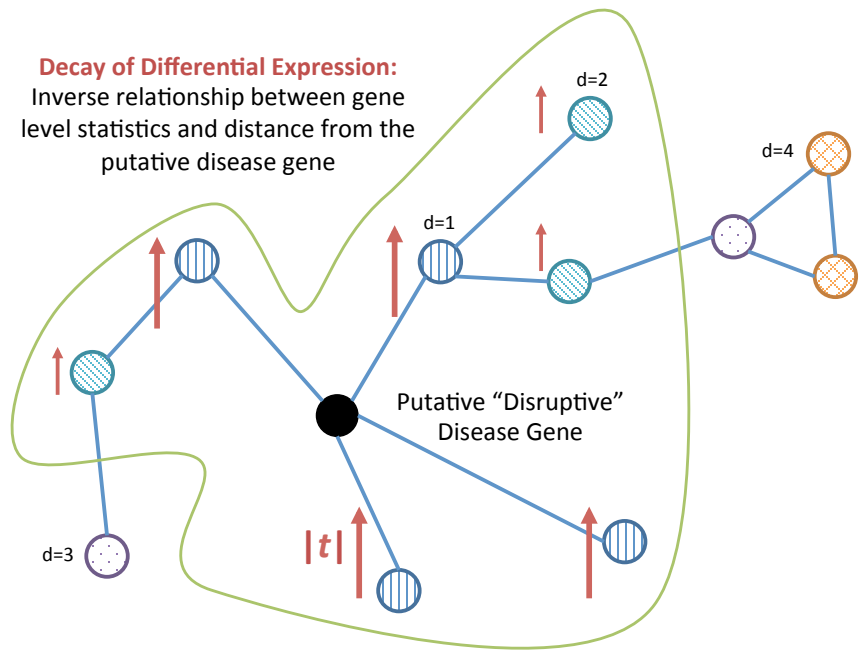}

\caption{\textbf{Procedure for Decay of 
Differential Expression.} The Decay of Differential Expression
computation tests if the discordance between differential expression
and distance from the driver gene is greater with the phenotype labels
we observe than with a random permutation of the sample labels.}

\label{fig:f2} 
\end{figure}

\comment{
The first test, \textit{Sphere of Influence},
calculates the evidence that a putative disease gene is correlated with its
neighbors. We compute the total correlation in its neighborhood and model it as
drawn from a random sample of genes
(Figure~\ref{fig:f1})
The second test, \textit{Decay of Differential Expression},
calculates the evidence that the neighbors of a putative disease gene are
differentially expressed (with the magnitude of differential expression
decreasing with distance). We quantify this pattern of differential expression
by calculating the discordance between differential expression  and distance
from the putative gene and model the differential expression  as drawn from a
random permutation of the phenotype labels 
(Figure~\ref{fig:f2}).
}

\subsection*{Does a gene appear to influence its neighbors in the network? Evidence of ``Sphere of Influence''}

If a gene is a source of regulatory control or disruption, we may expect to see that
its behavior is correlated with that of its neighbors.
The first step, dubbed ``Sphere of Influence,'' assesses if a candidate gene $i$
meets this criterion by testing if gene $i$ is more strongly correlated
with its network neighbors than would be expected by chance (Figure~\ref{fig:f1}),
compared to a random sample of genes.
The first step, therefore, of the Sphere of Influence procedure is to calculate
the  Spearman rank correlation $\rho_{ij}$  between gene $i$ and
every other gene $j$ assayed and on the network.
From this set of correlations, we calculate the observed total (absolute) correlation between gene $i$ and its neighbors within a neighborhood of radius $r$, 
\begin{equation}
C_i(r) = \sum_{\{j:d_{ij}{\leq}r\}} |\rho_{ij}| \,,
\label{eq:totalcorRB}
\end{equation}
where $d_{ij}$ indicates the geodesic distance of gene $j$ from gene $i$ on the
network.

In order to compute the distribution of total correlation under the null
hypothesis that it is drawn from a random sample of genes, we re-sample with
replacement from the set of correlations between gene $i$ and every other gene
$j$ and recompute Equation~\ref{eq:totalcorRB}. This procedure effectively
redistributes the gene--gene correlations about the network, enabling a
comparison of gene $i$'s influence in the true network neighborhood to its
influence on a random  selection of genes. This step tests the so--called
``competitive null''  described in~\citep{Khatri2012}; that is, whether gene $i$
has a greater correlation with genes in its neighborhood than would be a
expected from a random set of genes.

The null distribution of the total absolute correlation for gene $i$ as a
function of the neighborhood radius is computed using $10^3$ re-samplings, and
the observed total absolute correlation is compared to the re-sampled  null
distribution, yielding a ``Sphere of Influence'' $p$--value  at each  neighborhood radius for gene $i$,
$p^{\text{S}}_i(r)$, that quantifies whether $i$ is more correlated with its
neighbors than expected by chance, evidence that it may be an influential gene.

 \subsection*{Does the gene's neighborhood exhibit an association with phenotype? Evidence of ``Decay of Differential Expression''}

The previous step tests whether gene $i$ is strongly correlated with its network 
neighbors, independent of phenotype. 
If a gene is a source of disease-associated disruption, we may expect that it
and its neighbors will exhibit differential expression. 
We thus now turn our attention to whether
the gene and its neighbors also exhibit an association with the phenotype of interest.
In particular, if a gene $i$ is a source of dysregulation that drives 
the phenotype, we would expect that gene $i$ and its close neighbors will be 
differentially expressed, while genes farther away in the network will exhibit 
weaker differential expression. In other words, we expect a pattern of
differential expression that is strongly localized about $i$ and decays as 
one moves farther from it in the network. 
Hence, the second calculation, ``Decay of Differential Expression,'' tests whether
the magnitude of differential expression of other genes $j$ in the neighborhood
is inversely related to the distance $d_{ij}$ of gene $j$ from gene $i$
(Figure~\ref{fig:f2}).

In order to do this, we must first compute a gene--level statistic $g_j$ that
quantifies the magnitude of $j$'s association with the outcome of interest. We
then quantify the ``decay of differential expression'' with the Kendall
$\tau_{B}$ rank correlation  coefficient between the differential expression and
distance from gene $i$.

The observed discordance is 
\begin{equation}
    D_i(r) = \tau_{B}\left( \{g_j:d_{ij}{\leq}r\}, \{d_{ij}:d_{ij}{\leq}r\} \right)\,,
\label{eq:discordance}
\end{equation}
where $d_{ij}$ is the geodesic distance between gene $j$ and gene $i$.

To assess the statistical significance of $D_i(r)$, we randomly permute the 
phenotype labels and recompute the gene--level association statistics $g_j$
under the null hypothesis that the genes are not meaningfully associated with
the phenotype.  We then use the permuted $g^*_j$ to recompute $D^*_i$ according
to Equation~\ref{eq:discordance}.  A set of $10^3$ such re-computations forms
a reference distribution against which we compare the observed $D_i$ to obtain
a $p$ value $p^{\text{D}}_i(r)$ as the fraction of $D^*_i<D_i$.


It should be noted here that while $p^{\text{S}}_i(r)$ (above) was obtained
by randomly permuting \textit{genes}, $p^{\text{D}}_i(r)$ is obtained by
permuting the class labels.  An important feature of the latter is that it
preserves correlations between genes that were found in the
$p^{\text{S}}_i(r)$ calculation.  In consequence, the null models, and
hence the interpretations, of the two tests differ. $p^{\text{S}}_i(r)$
quantifies whether the neighborhood surrounding gene $i$ is more strongly
correlated with it than a random set of genes would be (independent of
phenotype), testing the so--called ``competitive null''~\citep{Khatri2012}.  In
contrast, $p^{\text{D}}_i(r)$ assesses whether the neighborhood surrounding
gene $i$ is more strongly associated with the phenotype of interest than those
same genes would be with randomly--assigned phenotype labels (preserving the
organization of genes in the network), thus testing the so--called 
``self-contained null''~\citep{Khatri2012}. That is, it tests whether a specific set of
genes in a neighborhood is more strongly associated with the phenotype of
interest than the same set of genes would be for a random phenotype.

Because these two procedures permute orthogonal axes (genes \vs samples), they
provide two independent tests with independent interpretations:
$p^{\text{S}}_i(r)$ tests whether gene $i$ influences its neighbors, and
$p^{\text{D}}_i(r)$ tests whether that neighborhood is associated with
disease.  We then combine these independent pieces of evidence into
a single assessment, as described below.

\subsection*{Combined Evidence} 

At this point in our algorithm, the Sphere of Influence and Decay of
Differential Expression procedures have been run independently of each other,
but neither component is sufficient by itself to determine if putative disease
gene $i$ is in fact a ``disruptive'' candidate disease gene meeting both criteria. 
Therefore, the last step our method performs is to combine the $p$-values
outputted by each component ($p^{\text{S}}_i(r)$ and $p^{\text{D}}_i(r)$)
using Fisher's method~\citep{Fisher1932-rp},
\begin{equation}
    X^2 = -2(\ln(p^{\text{S}}_i(r))+\ln(p^{\text{D}}_i(r)))\,.
\label{eq:pcomb}
\end{equation}
$X^2$ follows a $\chi^2$ distribution with 4 degrees of freedom, which
can be used compute $p^{\text{Comb}}_i(r)$, the combined evidence that gene $i$
is a ``disruptive'' gene.

\subsection*{Neighborhood Size} 

Above we described the Sphere of Influence and Decay of Differential Expression
procedures for a fixed radius ($r$), but different genes may have different
extents of influence on the network, and this extent is not known \textit{a priori}.
Therefore, we have devised our analysis technique  to apply the  Sphere of
Influence, Decay of Differential Expression, and Combined Evidence calculations
to the neighborhood of every radius (up to $D$ the diameter of the network).
The $p$-value our method outputs for each gene ($p^{\text{GS}}_i$),  therefore,
is the smallest $p^{\text{Comb}}_i(r)$ across all distances.

To evaluate the significance of $p^{\text{GS}}_i$, we then apply a
Bonferroni correction to the significance threshold  to conservatively adjust
for the multiple hypothesis tests that we perform when applying our method to
the neighborhoods of each radius. Since the number of  neighborhoods (and
therefore number of tests) is determined by the diameter of the network, we
scale the significance threshold by the diameter of the network to
determine whether a gene was significantly found to be ``disruptive'' in the
data.
Adjustment for the multiplicity of genes tested is achieved through
permutation as previously
described~\cite{camargo2008permutation,dudoit2003multiple,Subramanian2005};
this has the important benefit of preserving the biologically-relevant
dependency structure between
genes~\cite{dudoit2003multiple,Subramanian2005}.

\subsection*{Example of GeneSurrounder steps applied to an example gene}

\begin{figure}[hbt]
\centering
\includegraphics[width=\columnwidth]{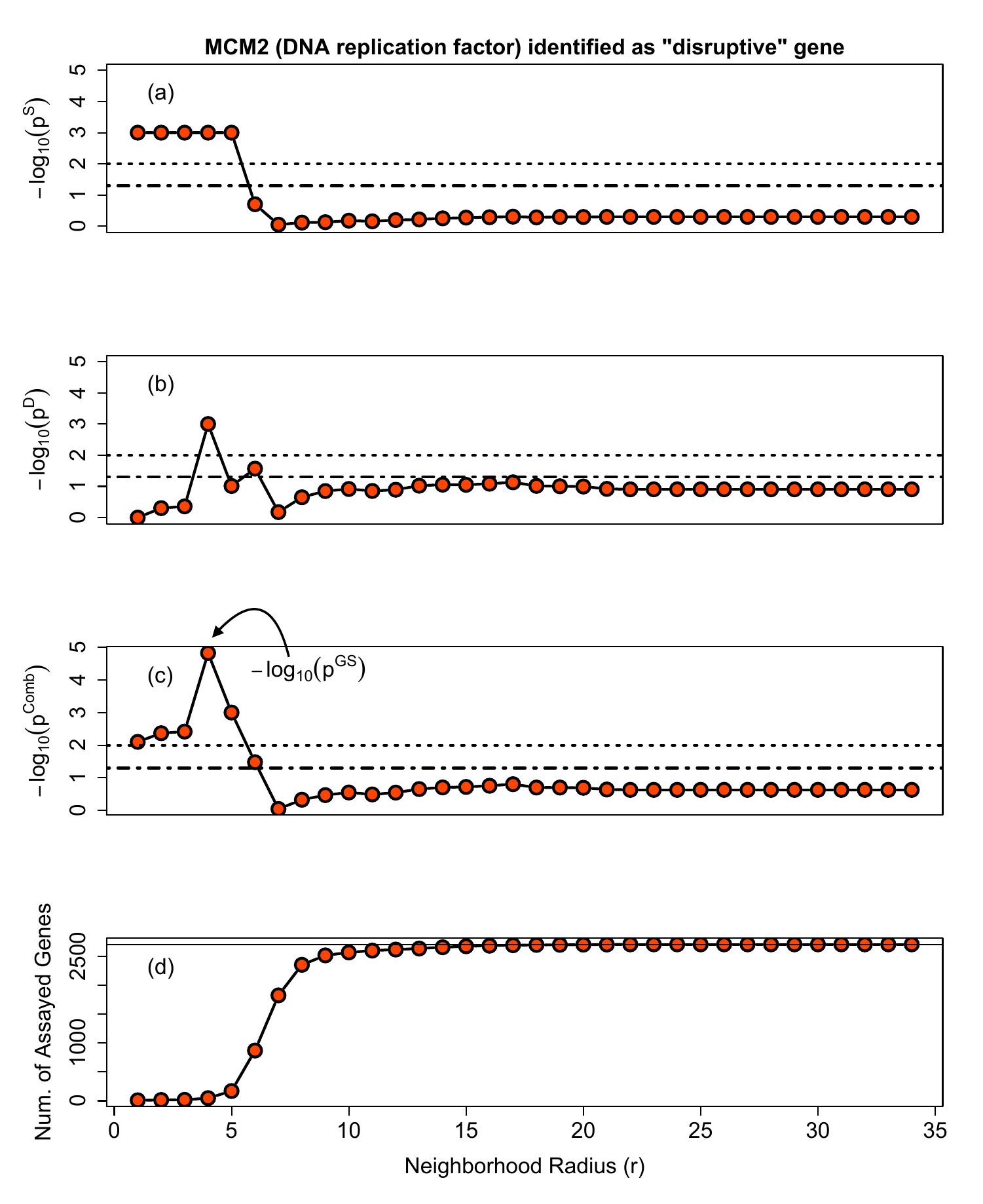}

\caption{\textbf{Illustration of Method}. Displayed are the results for 
the gene MCM2 when our algorithm was applied to Ovarian Cancer Study 
GSE14764. (a) shows ${-}\log_{10}(p_{\text{Sphere}})$ vs the Neighborhood
Radius. (b) shows ${-}\log_{10}(p_{\text{Decay}})$ vs the Neighborhood
Radius. (c) shows ${-}\log_{10}(p_{\text{Combined}})$ vs the Neighborhood
Radius. (d) shows the Number of Assayed Genes vs the Neighborhood
Radius. In the top three plots, the dashed and dotted lines correspond to 
a significance level of 0.05 and 0.01 respectively. In
the bottom plot, the solid line corresponds to the total
number of genes assayed and on the network.} 


\label{fig:illus} 
\end{figure}

To illustrate the components of the GeneSurrounder computation, we present the results for each component of our algorithm as
applied to gene \textit{MCM2} using data from one study of high-vs-low grade ovarian
cancer~\citep{Ganzfried2013} (GEO accession GSE14764). In Figure~\ref{fig:illus},
each of the first three plots (from top to bottom) displays the
$-\log_{10}(p)$ from the Sphere of Influence, Decay of Differential Expression
and Combined components of our method. Since we compute these values as a
function of network neighborhood size surrounding that gene, the $p$-values are
plotted against the neighborhood radius (\ie radius of geodesic distance from the
putative ``disruptive'' disease disease gene \textit{MCM2}.)

Figure~\ref{fig:illus}A (Sphere of Influence) illustrates the dilution of
influence with distance and the effect that the size (\ie number of assayed
genes) of a neighborhood has on the decrease of influence. The putative disease
gene in this example, \textit{MCM2}, has significant influence in neighborhoods near to
it, but this influence falls off and stays non-significant at far-away
distances. The largest difference occurs between a radius of 5 and 6,
where the number of assayed genes within the neighborhood (Figure~\ref{fig:illus}D)
increases sharply, contributing to the dilution of \textit{MCM2}'s influence.

Figure~\ref{fig:illus}B (Decay of Differential Expression) indicates a significant
concentration of differential expression for neighborhoods with radii of 4--6. 
We observe that small neighborhoods immediately near a putative disease gene are not big enough to detect
a decaying pattern of differential expression, such that the localized differential expression is only detectable at with a radius of at least 4. At the other end, big neighborhoods are too diverse to
exhibit a consistent decay of differential expression; like the sphere of influence,
the significance of the decay of differential expression flattens out at large distances. 

Figure~\ref{fig:illus}C illustrates the results of combining the results for
each neighborhood. The $p$-value our method outputs for each gene is the most
significant $p^{\text{Comb}}_i(r)$ across all neighborhood radii; for \textit{MCM2}
in this study, this occurs at a neighborhood radius of 4 with $p_\mathrm{GS}=1.48\e{-05}$. 
Since our
method returns the smallest $p^{\text{Comb}}_i(r)$ for each gene (equivalently,
the largest $-\log_{10} p^{\text{Comb}}_i(r)$)
and the smallest $p^{\text{Comb}}_i(r)$ of \textit{MCM2} is highly
significant, \textit{MCM2} would be identified as a central candidate disease gene
of high grade ovarian cancer. From a biological standpoint, this finding is
sensible: \textit{MCM2} is a DNA replication factor, and therefore likely plays a role
in the aggressive proliferation associated with high-grade ovarian carcinoma.

\section{Results} 

\subsection*{Application to Ovarian Cancer Data with Global KEGG Network Model}

We apply our algorithm to three gene expression data sets of high-vs-low grade
ovarian cancer from the publicly available and curated collection
`curatedOvarianData'~\citep{Ganzfried2013} to illustrate the components of the
GeneSurrounder method and evaluate its performance. In order to test our
algorithm, we evaluate its cross-study concordance, \ie, its consistency across
different data sets that are measuring the same conditions, as previously
described~\citep{SHAH2014}. The intuition underlying this approach is that
methods that detect true biological signals should find them across different
data sets measuring the same conditions. To test this, we use data from three
independent studies of gene expression of $7680$ genes (common in all three
studies) in patients with high and low grade cancer (Table~\ref{tab:ovdattab}).
The data was downloaded using the Bioconductor package 
`curatedOvarianData'~\cite{Ganzfried2013}.
The curatedOvarianData project was designed to facilitate
meta-analysis and therefore provides data that has 
been harmonized to ensure that clinical data (such as grade) are comparable
across studies. Gene expression data was preprocessed by the original authors
using established normalization techniques, and no further preprocessing was
required

Our method combines two independent sources of information --- the  gene
expression data and a pathway network model --- to detect the disruptive genes
of the phenotype under consideration. We use the same global network model for
each study, which we have constructed from KEGG pathways~\citep{Kanehisa2008}.
The KEGG database organizes experimentally derived  pathway information into
individual networks of functionally related molecules. In the KEGG
representation, the nodes (\ie vertices) are genes or gene products, and the
links (\ie edges) are cellular interactions.  We create a global KEGG network to
avoid the artificial boundaries between individual pathways by
taking the graph union of the individual pathways, \ie merging the
pathways so that the links which are in at least one KEGG pathway will be part
of the new global network. We then consider the largest connected component of
the resulting network in our algorithm. This global network has $N=4867$ nodes,
$L=42874$ edges, and a diameter $D=34$. Of the $N=4867$ nodes, $2709$ of them
are also amongst  the $7680$ genes assayed in all three ovarian cancer studies.

We apply our method to each of the ovarian cancer studies with the global gene
network to calculate the combined evidence $p^{\text{Comb}}_i(r)$
for each of the $2709$ genes $i$ that are assayed and on the network.
A table of the full results is provided as an additional file [see
Additional file 1, Additional file 2, Additional file 3].
With the
results from each of the three ovarian cancer data sets, we evaluate not only
the cross-study concordance of our analysis technique, but also its ability to
identify biologically relevant  genes and truly integrate pathway and expression
data.

\subsection*{Disruptive genes found by GeneSurrounder are associated with
ovarian cancer}

To evaluate GeneSurrounder's ability to identify biologically relevant genes, we
compare our results in all three ovarian cancer studies
(Table~\ref{tab:ovarian_threshold}) to existing biological knowledge. Applying
GeneSurrounder to the $2709$ common genes between studies that were assayed and
on the network, we generated three distinct ranked lists of genes for each study
based on the computed $p^{\text{GS}}_i$ value. To compare these results to
existing biological knowledge, we consider genes that pass our Bonferroni
corrected threshold (at significance level $\alpha=0.05$
and with a  diameter of $D = 34$, our Bonferroni corrected threshold is 
$-\log_{10}(p) \geq 2.83$) in all
three studies (Table~\ref{tab:ovarian_threshold}).

We used the DOSE R package~\cite{Yu2015-ui} to analyze the enrichment of these
genes with Disease Ontology (DO) terms~\cite{Schriml2012-zj}.  We found that the
genes that pass our Bonferroni corrected threshold in at least one  ovarian cancer study
were significantly enriched  with the DO term ``ovarian cancer''
(DOID:2394) ($p = 0.0000177$). Furthermore, amongst these genes are three families
of protein coding genes,  \textit{CDC} (involved in the cell division cycle),
\textit{MCM}, and \textit{ORC} (both involved in DNA replication), with
biological  functions that support their role in ovarian cancer.

To further compare our results to existing biological knowledge,  we  found
evidence in the literature that \textit{CDC7}, \textit{ORC6L}, and \textit{DBF4}
are associated specifically with ovarian
cancer~\citep{Kulkarni2009-xh,Bowen2009-py,Bonte2008-of}. The inclusion of
\textit{CDC45} suggests the possibility that it is also associated with ovarian
cancer. \textit{CDC7} encodes for a cell division cycle protein and has been
found to both predict survival and be a powerful anticancer target in ovarian
cancer~\citep{Kulkarni2009-xh}. \textit{ORC6L} encodes for a origin recognition
complex that is crucial for the initiation of DNA replication and has been found
to highly expressed in ovarian cancer~\citep{Bowen2009-py}. \textit{DBF4}
encodes for a protein that activates the kinase activity of \textit{CDC7} and
was found to be associated with ovarian cancer~\citep{Bonte2008-of}. The finding
of these  genes from studies of high-vs-low grade ovarian cancers suggests the
possibility that they are not only involved in ovarian cancer but, more
specifically, drive high grade ovarian cancer.  A table of the full results is
provided as an additional file [see Additional file 1, Additional file 2,
Additional file 3].

\subsection*{GeneSurrounder results represent a true integration of pathway and expression data}

The method that we have developed combines gene expression data with an
independent network model. To investigate whether our results are driven solely
by either the network or the expression data or represent a true integration
of biological knowledge (the pathway networks) and experimental data, we
consider the association between our results, the centrality, and the
differential expression for each gene. If the results were driven solely by the
network, the evidence a gene is a disruptive gene would correlate strongly with
its centrality in the network. We therefore calculate the correlation between
our results and two different measures of centrality. If the results were driven
solely by the expression data, the evidence a gene is a disruptive gene
would correlate strongly with its differential expression 
We therefore calculate the correlation between our results and the differential
expression for each of the studies. The
results are given in Table~\ref{tab:centrality}. We find that for each of the
studies, the correlations are small (on the order of $0.01$), confirming that
GeneSurrounder is not driven solely by network features or the
expression data, but rather represents a true integration of biological
knowledge (the pathway networks) with experimental data.

\subsection*{GeneSurrounder findings are more concordant than differential
expression analysis}

The intuition underlying evaluating cross-study concordance is that methods
that detect true biological signals should find them across different data sets
measuring the same conditions.
To investigate the cross-study concordance of our analysis technique (\ie its
consistency across different data sets measuring the same conditions),
we consider each pair of the three studies and calculate the correlation
between our results.  
As a point of reference, we also calculate the
correlation between the gene level statistics obtained using the customary 
$t$-test for differential expression. The
results are given in Table~\ref{tab:xstudy}.
As mentioned earlier, methods that do not take into account systems-level
information tend to have poor agreement between studies because the individual genes
contributing to disease-associated mechanisms can vary from one study to the next.
Indeed, we find that the cross-study concordance of differential expression 
results is remarkably low (Table~\ref{tab:xstudy}). 
By contrast our method is 3.51---8.55 times more consistent than differential 
expression analysis.
This cross--study concordance suggests that our method reliably detects
biological effects reproducibly across studies.

\subsection*{GeneSurrounder findings are more concordant than LEAN}

We also compare GeneSurrounder to LEAN, a recent method that also attempts to
integrate gene expression and network data to identify significant genes.
In contrast to our method, LEAN considers only the immediate neighborhood (\ie at a
radius of one) and assesses the enrichment of significant genes.
To compare the performance of our analysis technique to LEAN,
we compare their respective cross-study concordances. To ensure comparability 
between our method and LEAN, we use the same network and expression data for
inputs to LEAN that we used for GeneSurrounder. 
Again, we 
consider each pair of the three studies and calculate the correlation
between our results and the correlation between results of
LEAN~\citep{Gwinner2016-kq} (which is available as an R package on CRAN). The
results are given in Table~\ref{tab:xstudy}.
%
We found that while LEAN
is more consistent than the differential expression analysis, GeneSurrounder is more
consistent than LEAN.
That is, the list of ``disruptive'' genes
detected by GeneSurrounder are more reproducible across studies than
both differentially expressed genes and the results from LEAN.

\subsection*{Application to Bladder Cancer Data with Global KEGG Network Model}

As a further demonstration of our method, we apply our algorithm to three gene expression data sets of superficial-vs-invasive bladder cancer from the publicly available and curated collection
`curatedBladderData'~\cite{Riester2012-vi}. As in our application to ovarian cancer, the bladder cancer data was downloaded using the
Bioconductor package `curatedBladderData' and we did not do any additional 
pre-processing to the bladder cancer datasets used in this study. The same global
KEGG network we created to analyze the Ovarian Cancer Data was used to analyze
the Bladder Cancer Data. Of the $N=4867$ genes in our global KEGG network,
$3406$ of them were common to all three bladder cancer studies
(Table S1) and  could be mapped to KEGG IDs using the
org.HS.eg.db package~\cite{Carlson2018}. We apply our method to each 
of the the $3406$
genes that were in the microarray data and on the network for each bladder
cancer study and 
evaluate the cross study concordance in these additional datasets, as well as GeneSurrounder's ability to identify known bladder cancer-associated genes.
A table of the full results is provided as an additional file [see Additional
file 4, Additional file 5, Additional file 6].

We find that while the results of our method ($p^{\text{GS}}$) for GSE13507 and
GSE32894 are similar to their expected theoretical distributions (the  minimum
of $N=34$ independent $U(0,1)$ random variables) the results for GSE31684 do not.
%
We expect that this is due to the comparatively low sample size of GSE31684, which had only 15 samples in the "superficial" classification.
As a result, when finding genes that pass our Bonferroni corrected
threshold (at a threshold of $p = 0.05$ and with a  diameter of $D = 34$, our
Bonferroni corrected threshold is $-\log_{10}(p) \geq 2.83$) none of the genes
in GSE31684 pass this threshold. However, we do find genes that pass this
threshold in both GSE13507 and GSE32894 (Table S2).
We also confirm that GeneSurrounder is not driven solely by network features
or the bladder cancer expression data (Table S3).
While all concordance values for all methods were generally lower in the bladder cancer studies than in the ovarian cancer studies, we nevertheless find that GeneSurrounder is still more concordant than both
differential expression analysis and LEAN (Table S4). 
Complete results of the bladder cancer data analysis, including discussion of significant genes, is provided in a Supplementary file [See Additional file 7].

\section{Discussion} 

In this manuscript, we have developed and presented a new analysis technique,
GeneSurrounder, that integrates a network model with expression data to identify
individual genes that can be targeted therapeutically. 
Our analysis technique identifies ``disruptive'' genes --- genes that impact
pathway networks in a disease associated manner. 
The algorithm consists of two tests that are run independently of each
other and then combined. The first test, Sphere of Influence,
calculates the evidence that a putative disease gene is correlated with its
neighbors, and the second test, Decay of Differential Expression, calculates the
evidence that the neighbors of a putative disease gene are differentially
expressed (with the magnitude of differential expression decreasing with
distance).

We applied our algorithm to three gene expression data sets of high-vs-low grade
ovarian cancer~\citep{Ganzfried2013} and combined each of them with the same
global network model that we constructed from KEGG pathways. With the results
from each of the three ovarian cancer data sets, we evaluated our analysis
technique. By applying our method to three different data sets
measuring the same conditions, we were able to show that it yields consistent
(\ie concordant) results across studies, suggesting its ability to detect
biologically meaningful associations that are reproducible across studies. We
also compare our results to existing biological knowledge and find that our
method identifies biologically relevant genes.  To show that our method
truly integrates
pathway and expression data, we compare the  results from our method to the
results from a single gene analysis and  the centrality of the genes in the
network. Our positive results along these three dimensions of our analysis
technique suggest that our method is a promising new strategy for identifying
the genes that control disease.

As discussed in the Introduction, pathway analysis techniques such as
GSEA~\citep{Subramanian2005}, jActiveModules~\citep{Ideker2002}, and
COSINE~\citep{Ma2011-ag} use interaction networks and expression data to find
groups of related disease-associated genes. GeneSurrounder, to make experimental
follow-up easier, identifies precise gene targets rather than groups of tens or
hundreds of genes.  Efforts to identify individual genes, as our method does,
have either required prior biological knowledge (as in
ENDEAVOUR~\citep{Aerts2006-in} and GeneWanderer~\citep{Kohler2008-nr}) or have
not used direct interactions on a global network (as in~\citep{Nitsch2009}, an
extension of SPIA~\citep{Shafi2015-qu}, and nDGE~\citep{Wu2013-tp}).  Our
analysis technique addresses these shortcomings by using the shortest  direct
distance on a global network and not requiring any prior biological knowledge.
LEAN~\citep{Gwinner2016-kq} considers interactions on a global interaction
network and is closest to our method in this respect, but restricts its focus to
nearest neighbors on the network and does not determine  whether a putative
disease gene is the source of change on the network. 

\section{Conclusions} 

The key innovation of
GeneSurrounder is the combination of  pathway network information with gene
expression data to determine the degree to which a gene is a source of
dysregulation on the network.
GeneSurrounder employs a novel strategy by finding genes that both appear to
influence nearby genes and cause dysregulation associated with the disease.
Because GeneSurrounder considers every neighborhood size around a putative gene,
it is able to identify disease genes that may have broad effects on the regulatory
network (beyond nearest neighbors). 
GeneSurrounder thus provides a new avenue for identifying disease-associated genes
by detecting genes that appear to be sources of change and could therefore be
promising therapeutic targets.

While our method performs well in practice, there are limitations that bear consideration.
We note that the the network model that we use, KEGG, is not   
phenotype-specific (as are most pathway databases) and we therefore have to assume that
the network does not change between conditions. 
Additionally, because KEGG (and other pathway databases) may not be complete, genes which are not known to play a role any pathway cannot be considered in a GeneSurrounder analysis.
Furthermore, as implemented our algorithm
calculates geodesic distances between genes without taking into account the
direction or type of interactions. 
However, we note that our approach as presented here could easily be modified 
to take in as input
other kinds of networks (including context-specific computationally derived
networks) and/or considering edge directionality by changing the gene-gene 
distance matrix that the  Sphere of Influence and Decay of Differential 
computations use.

GeneSurrounder
can be potentially generalized to other types of data. For
instance, one might envision applying it to other kinds of omic data. For
example, GeneSurrounder could potentially be extended to use genomic sequence
data to identify epistatic interactions, evidenced by gene neighborhoods that
have a high level of correlations in their genetic variants. Our
method could also possibly be generalized for time-series gene expression data 
by either changing the gene-level statistics used by the algorithm or applying
it separately to time points.

GeneSurrounder thus provides means to prioritize genes that are sources of
disruption for a disease in the context of gene regulatory networks. By
prioritizing genes in this way, our method provides insights into disease
mechanisms and suggests diagnostic and therapeutic targets. Our method can be
used to help biologists select among tens or hundreds of genes for further
validation. Furthermore, it can be generalized to other kinds of networks
(including context-specific  networks) and omic data. This approach can not only
be used directly to prioritize promising targets, but also suggests new
strategies for integrating systems level information with omic data to identify,
validate, and target  disease mechanisms. We have made the implementation of our
method available to researchers on GitHub at \url{http://github.com/sahildshah1/gene-surrounder} 
with the aim of furthering our  understanding of statistical
techniques to identify disease-associated genes.

\clearpage
\section*{Declarations} 

\subsection*{Ethics approval and consent to participate}

Not applicable

\subsection*{Consent to publish}

Not applicable

\subsection*{Availability of data and material}



The expression datasets analyzed during the current study are available as 
part of the curatedOvarianData~\cite{Ganzfried2013} package and curatedBladderData package~\cite{Riester2012-vi}
in the Bioconductor repository,
https://bioconductor.org/packages/release/data/experiment/html/curatedOvarianData.html and
https://bioconductor.org/packages/release/data/experiment/html/curatedBladderData.html.
The network datasets analyzed during the current study are available as part of the 
KEGG repository, \url{http://www.genome.jp/kegg/}
via the KEGGgraph package~\cite{Zhang2009}.
All data generated during this study are included in this published article 
as supplementary information files.
An R implementation of the GeneSurrounder method is available from
\url{http://github.com/sahildshah1/gene-surrounder}

\subsection*{Competing interests} 

The authors declare that they have no competing interests.

\subsection*{Funding} 

This work was supported by the James S.\,
McDonnell Foundation (grant \#220020394), the National Heart, Lung, and Blood
Institute of the National Institutes of Health under award number R01HL128173,
and the  Fishel Fellowship in Cancer Research.
The content is solely the responsibility of the authors and does not necessarily
represent the official views of the National Institutes of Health.

\subsection*{Authors' contributions} 

SDS and RB conceived and designed the study. SDS developed the method, implemented it in R,
applied it to the data, and analyzed the results.  SDS and RB wrote the manuscript.
All authors read and approved the final manuscript.

\subsection*{Acknowledgments} 

We would like to thank Dr. William Kath, Dr. Marta Iwanaszko, Dr. Gary Wilk,
Dr. Phan Nguyen, Elan Ness-Cohn, Eric Johnson, Caitlin H. Garvey, Dr. Seth Corey,
and Dr. Marek Kimmel for helpful discussions.

\clearpage
\section*{Tables}






\setlength{\tabcolsep}{15pt} 



\begin{table}[!hb]
\centering

\begin{tabular*}{0.75\textwidth}{ l  c  c }
\toprule
GEO Accession No. & $N$(low-grade) & $N$(high-grade) \\ 
\midrule
  GSE14764 &  24 &  44 \\ 
  GSE17260 &  67 &  43 \\ 
  GSE9891 & 103 & 154 \\ 
\bottomrule
\end{tabular*}

\caption{\textbf{Ovarian cancer datasets used in this study}: Comparisons were made between low-- and high--grade serous
ovarian carcinoma using public data.  Sample sizes for each group 
in each dataset are given.  The data are publicly accessible and
available as part of the curatedOvarianData package~\citep{Ganzfried2013}.}
\label{tab:ovdattab}

\end{table}



\setlength{\tabcolsep}{15 pt} 

\begin{table}[!hb]

\centering


\begin{tabular}{l c c c}
  \toprule
     & \multicolumn{3}{c}{$-\log_{10}p^{\text{GS}}$}\\
     \cmidrule{2-4}
      Gene & GSE14764 & GSE17260  & GSE9891 \\
     \midrule

    \textit{ADRB3}  & 3.033 & 2.933 & 3.554 \\
    \textit{AURKA}  & 2.865 & 3.383 & 3.716 \\
    \textit{CDC45}  & 4.270 & 3.741 & 4.830 \\
    \textit{CDC7 } & 4.386 & 3.769 & 4.830 \\
    \textit{DBF4 } & 4.270 & 3.769 & 4.830 \\
    \textit{IL7 }  & 3.055 & 2.898 & 2.910 \\
    \textit{ITGAM}  & 2.961 & 3.024 & 3.094 \\
    \textit{MCM2 } & 4.830 & 3.372 & 4.830 \\
    \textit{MCM3 } & 4.830 & 3.383 & 4.830 \\
    \textit{MCM4 } & 4.830 & 3.394 & 4.830 \\
    \textit{MCM5 } & 4.830 & 3.372 & 4.830 \\
    \textit{MCM6 } & 4.830 & 3.428 & 4.830 \\
    \textit{ORC4 } & 4.386 & 3.172 & 4.830 \\
    \textit{ORC6 } & 4.386 & 3.691 & 4.830 \\
    \textit{TTK }  & 2.904 & 3.089 & 4.830 \\

  \bottomrule
\end{tabular}

\caption{\textbf{``Disruptive'' disease genes in high-grade ovarian cancer
consistently found by GeneSurrounder}: At a threshold of $p=0.05$ and with a
diameter of $D=34$, the Bonferroni  corrected threshold is  $-\log_{10}(p) \geq
2.83.$ Listed are the genes that pass this threshold in all three studies.}

\label{tab:ovarian_threshold}

\end{table}



\setlength{\tabcolsep}{15pt} 



\begin{table}[!hb]


\centering

\begin{tabular*}{\columnwidth}{ l  c  c  c }%
  \toprule
  Network/Gene Statistic & GSE14764 & GSE17260 & GSE9891 \\
  \midrule
  Degree Cor. & 0.044 & 0.070 & 0.038  \\
  Betweenness Cor. & 0.047  & 0.059 & 0.030 \\
  $p_{\text{DE}}$ Cor. &  0.060  & 0.103  & $- 0.051$ \\
  \bottomrule
\end{tabular*}

\caption{\textbf{Correlation between GeneSurrounder results and network/gene statistics}: The three columns are the
rank correlation between GeneSurrounder results ($p^{\text{GS}}$)
and network/gene statistics (Degree, Betweenness, and $p^{\text{DE}}$)
across all genes in each dataset. The Degree and Betweenness are two
different network centrality measures. The Degree is the number of
connections a node has and the Betweenness is the fraction of 
shortest paths that passes through the node. 
$p^{\text{DE}}$ is the $p$-value obtained from a standard 
differential expression $t$-test.}

\label{tab:centrality}


\end{table}

\setlength{\tabcolsep}{15pt} 

\begin{table}[!hb]
\centering

\begin{tabular*}{\columnwidth}{ l  c  c  c }

  \toprule
  Ovarian Cancer Study Pair & $p^{\text{GS}}$ Cor. & $p^{\text{DE}}$ Cor. & $p^{\text{LEAN}}$ Cor. \\
  \midrule
  GSE14764 - GSE17260 & 0.342  & 0.040 & 0.056 \\ 
  GSE14764 - GSE9891 & 0.436 & 0.056 & 0.130 \\
  GSE17260 - GSE9891 & 0.485 & 0.138 & 0.290 \\
  \bottomrule
\end{tabular*}

\caption{\textbf{Cross study concordance of GeneSurrounder results compared to differential
expression analysis and LEAN}: The columns $p^{\text{GS}}$ Cor., $p^{\text{DE}}$ Cor., and $p^{\text{LEAN}}$ Cor.
are the Spearman rank correlations respectively  between the results obtained from 
GeneSurrounder, differential expression analysis, and LEAN for each study pair.
}
\label{tab:xstudy}

\end{table}

\clearpage
\section*{Figures}

\clearpage
\bibliographystyle{unsrt}


%
%

\bibliography{main}

\clearpage

\section*{Supplement}

\renewcommand{\thetable}{S\arabic{table}}
\setcounter{table}{0}

\subsection*{GeneSurrounder Analysis of curatedBladderData}

In this supplement, we detail the results of applying GeneSurrounder to data
from the curatedBladderData package. The analysis parallels that of the ovarian
cancer data described in the main manusript.

\subsection*{Disruptive genes found by GeneSurrounder are associated with
bladder cancer}

To evaluate GeneSurrounder's ability to identify biologically relevant genes, we
compare our results in all three bladder cancer studies
(Table~\ref{tab:bladder_threshold}) to existing biological knowledge. Applying
GeneSurrounder to the $3406$ common genes between studies that were assayed and
on the network, we generated three distinct ranked lists of genes for each study
based on the computed $p^{\text{GS}}_i$ value. To compare these results to
existing biological knowledge, we consider genes that pass our Bonferroni
corrected threshold (at significance level $\alpha=0.05$ and with a  diameter of
$D = 34$, our Bonferroni corrected threshold is $−log_{10}(p)  \geq 2.83$) in
all three studies (Table~\ref{tab:bladder_threshold}).

We used the DOSE R package~\cite{Yu2015-ui} to analyze the enrichment of these
genes with Disease Ontology (DO) terms~\cite{Schriml2012-zj}.  We found that the
genes that pass our Bonferroni corrected threshold in at least one bladder cancer
were significantly enriched with the DO term ``bladder cancer''
(DOID:11054) ($p = 0.00280$).  
Furthermore, our method found \textit{DUSP1} and
\textit{TNFRSF13B}. \textit{DUSP1}  plays an important role in the negative
regulation of cellular proliferation  and can make some solid tumors resistant
to therapy. \textit{TNFRSF13B} belongs to the tumor necrosis factor (TNF) ligand
family and plays roles in events such as cell survival, proliferation,
differentiation, and death.  The finding of these genes from studies of
superficial-vs-invasive  bladder is sensible and suggests that they may not only
be involved in cancer, but, more specifically, superficial-vs-invasive bladder
cancer.
A table of the full results is provided as an additional file [see Additional
file 4, Additional file 5, Additional file 6].

\subsection*{GeneSurrounder results represent a true integration of pathway and expression data}

The method that we have developed combines gene expression data with an
independent network model. To investigate whether our results are driven solely
by either the network or the expression data or represent a true integration of
biological knowledge (the pathway networks) and experimental data, we consider
the association between our results, the centrality, and the differential
expression for each gene. If the results were driven solely by the network, the
evidence a gene is a disruptive gene would correlate strongly with its
centrality in the network. We therefore calculate the correlation between our
results and two different measures of centrality. If the results were driven
solely by the expression data, the evidence a gene is a disruptive gene would
correlate strongly with its differential expression  We therefore calculate the
correlation between our results and the differential expression for each of the
studies. The results are given in Table~\ref{tab:bladder_integrationdata}. We
find that for each of the studies, the correlations are small (at most
$+0.194$), confirming that GeneSurrounder is not driven solely by network
features or the expression data, but rather represents a true integration of
biological knowledge (the pathway networks) with experimental data.

\subsection*{GeneSurrounder findings are more concordant than differential
expression analysis}

The intuition underlying evaluating cross-study concordance is that methods that
detect true biological signals should find them across different data sets
measuring the same conditions. To investigate the cross-study concordance of our
analysis technique (\ie its consistency across different data sets measuring the
same conditions), we consider each pair of the three studies and calculate the
correlation between our results.   As a point of reference, we also calculate
the correlation between the gene level statistics obtained using the customary
$t$-test for differential expression. The results are given in
Table~\ref{tab:bladder_xstudy}. As mentioned earlier, methods that do not take
into account systems-level information tend to have poor agreement between
studies because the individual genes contributing to disease-associated
mechanisms can vary from one study to the next. Indeed, we find that the 
cross-study concordance of differential expression  results is remarkably low
(Table~\ref{tab:bladder_xstudy}).  By contrast our method is  more consistent
than differential  expression analysis. This cross--study concordance suggests
that our method reliably detects biological effects reproducibly across studies.

\subsection*{GeneSurrounder findings are more concordant than LEAN}

We also compare GeneSurrounder to LEAN, a recent method that also attempts to
integrate gene expression and network data to identify significant genes. In
contrast to our method, LEAN considers only the immediate neighborhood (\ie at a
radius of one) and assesses the enrichment of significant genes. To compare the
performance of our analysis technique to LEAN, we compare their respective
cross-study concordances. To ensure comparability  between our method and LEAN,
we use the same network and expression data for inputs to LEAN that we used for
GeneSurrounder.  Again, we  consider each pair of the three studies and
calculate the correlation between our results and the correlation between
results of LEAN~\citep{Gwinner2016-kq} (which is available as an R package on
CRAN). The results are given in Table~\ref{tab:bladder_xstudy}. We found that
GeneSurrounder is more consistent than LEAN. That is, the list of ``disruptive''
genes detected by GeneSurrounder are more reproducible across studies than both
differentially expressed genes and the results from LEAN.


\subsection*{Tables}

\setlength{\tabcolsep}{15pt} 

\begin{table}[!hb]
\centering

\begin{tabular*}{0.75\textwidth}{ l  c  c }

\toprule
GEO Accession No. & $N$(superficial) & $N$(invasive) \\ 
\midrule
  GSE13507 &  103 &  62 \\ 
  GSE31684 &  15 &  78 \\ 
  GSE32894 &  213 & 93 \\ 
\bottomrule
\end{tabular*}

\caption{\textbf{Bladder cancer datasets used in this study}:  Comparisons were
made between superficial and invasive  bladder cancer using public data.
Superficial bladder cancer has not grown into the main muscle layer of the
bladder and invasive bladder cancer has grown into the main muscle layer of the
bladder.  Sample sizes for each group  in each dataset are given.  The data are
publicly accessible and available as part of the curatedBladderData
package~\citep{Riester2012-vi}.}

\label{tab:bladder_datasets}

\end{table}

\setlength{\tabcolsep}{20 pt} 

\begin{table}[!hb]

\centering

\scriptsize{

\begin{tabular}{l c c c}
  \toprule
     & \multicolumn{3}{c}{$-\log_{10}p^{\text{GS}}$}\\
     \cmidrule{2-4}
      Gene & GSE13507 & GSE31684  & GSE32894 \\
     \midrule

    \textit{A2M} & 3.046 & 0.799 & 2.987 \\
    \textit{ABLIM2} & 3.992 & 0.415 & 4.107 \\
    \textit{CCL11} & 3.523 & 0.123 & 3.104 \\
    \textit{CCL19} & 4.107 & 0.121 & 2.980  \\
    \textit{CD14} & 4.046 & 0.281 & 3.416 \\
    \textit{CD209} & 3.296 & 0.229 & 2.895 \\
    \textit{CD22} & 3.452 & 0.166 & 3.016 \\
    \textit{CD226} & 3.383 & 0.257 & 2.950 \\
    \textit{CD247} & 3.479 & 0.195 & 3.041 \\
    \textit{CD3G} & 2.901 & 0.289 & 2.904 \\
    \textit{CD48} & 2.868 & 0.484 & 3.716 \\
    \textit{CSF2} & 4.830 & 0.028 & 3.416 \\
    \textit{CTSG} & 3.280 & 0.008 & 2.841 \\
    \textit{DOCK2} & 2.859 & 0.170 & 2.991 \\
    \textit{DUSP1} & 4.830 & 0.076 & 3.248 \\
    \textit{DUSP5} & 3.523 & 0.080 & 3.280 \\
    \textit{FYB} & 2.901 & 0.214 & 3.149 \\
    \textit{GRAP2} & 3.079 & 0.280 & 3.234 \\
    \textit{GZMA} & 3.903 & 0.210 & 2.901 \\
    \textit{IL18RAP} & 2.865 & 0.103 & 3.372 \\
    \textit{IL1A} & 3.120 & 0.107 & 2.991 \\
    \textit{IL1B} & 2.860 & 0.131 & 3.314 \\
    \textit{IL6} & 3.830 & 0.029 & 3.493 \\
    \textit{IQGAP2} & 2.995 & 0.284 & 3.691 \\
    \textit{IRF1} & 4.180 & 0.067 & 2.907 \\
    \textit{IRF8} & 3.428 & 0.960 & 3.264 \\
    \textit{ITGAM} & 4.046 & 0.412 & 3.173 \\
    \textit{JUN} & 2.947 & 0.134 & 2.991 \\
    \textit{LCP2} & 4.270 & 0.031 & 2.889 \\
    \textit{LIF} & 2.848 & 0.130 & 3.405 \\
    \textit{MAP3K8} & 3.669 & 0.107 & 3.342 \\
    \textit{MAP4K1} & 4.270 & 0.481 & 3.115 \\
    \textit{MEF2C} & 3.199 & 0.165 & 2.897 \\
    \textit{MSN} & 4.830 & 0.408 & 2.904 \\
    \textit{NCKAP1L} & 4.045 & 0.337 & 3.865 \\
    \textit{NKD1} & 3.669 & 0.100 & 3.830 \\
    \textit{NLRP3} & 3.522 & 0.120 & 2.901 \\
    \textit{NTF3} & 3.248 & 0.069 & 2.910 \\
    \textit{OSM} & 3.493 & 0.124 & 3.003 \\
    \textit{PDCD1LG2} & 3.865 & 0.112 & 2.995 \\
    \textit{PLAU} & 4.181 & 0.173 & 3.154  \\
    \textit{POMC} & 3.079 & 0.055 & 2.999 \\
    \textit{PTPN7} & 3.241 & 0.078 & 2.991 \\
    \textit{RAC2} & 3.160 & 1.008 & 3.589 \\
    \textit{RELB} & 4.829 & 0.422 & 4.386 \\
    \textit{RUNX1T1} & 3.234 & 0.118 & 2.987 \\
    \textit{SEC23B} & 3.104 & 2.102 & 3.041 \\
    \textit{SELP} & 4.830 & 0.153 & 3.226 \\
    \textit{SERPINA1} & 3.523 & 0.663 & 3.256 \\
    \textit{SERPING1} & 4.181 & 0.285 & 3.248 \\
    \textit{SH2D1A} & 3.992 & 0.245 & 3.992 \\
    \textit{SIRPG} & 3.323 & 0.027 & 2.857 \\
    \textit{SLIT2} & 3.945 & 0.125 & 3.185 \\
    \textit{TCL1A} & 2.954 & 0.458 & 3.093 \\
    \textit{THY1} & 3.296 & 0.080 & 2.910 \\
    \textit{TNFAIP3} & 3.016 & 0.213 & 3.305 \\
    \textit{TNFRSF13B} & 4.180 & 0.256 & 4.386 \\
    \textit{TNFSF13B} & 2.991 & 0.255 & 4.108 \\
    \textit{TYROBP} & 3.120 & 0.392 & 2.914 \\
    \textit{WAS} & 4.830 & 0.328 & 3.173 \\
    \textit{ZAP70} & 4.830 & 0.149 & 3.205 \\
    \textit{ZBP1} & 3.768 & 0.293 & 4.549 \\

\bottomrule
\end{tabular}

\caption{\textbf{``Disruptive'' disease genes in bladder cancer 
consistently found by GeneSurrounder}: At a threshold of $p=0.05$ and with a
diameter of $D=34$, the Bonferroni  corrected threshold is  $-\log_{10}(p) \geq
2.83.$ Listed are the genes that pass this threshold in GSE13507 and GSE32894.}

\label{tab:bladder_threshold}

}

\end{table}

\setlength{\tabcolsep}{15pt} 

\begin{table}[!hb]

\centering

\begin{tabular*}{\columnwidth}{ l  c  c  c }%
  \toprule
  Network/Gene Statistic & GSE13507 & GSE31684 & GSE32894 \\
  \midrule
  Degree Cor. & $+ 0.084$ & $+ 0.194$ & $+ 0.120$  \\
  Betweenness Cor. & $+ 0.063$  & $+ 0.084$ & $+ 0.058$ \\
  $p_{\text{DE}}$ Cor. &  $+ 0.107$  &  $+ 0.111$  & $- 0.016$ \\
  \bottomrule
\end{tabular*}

\caption{\textbf{Correlation between GeneSurrounder results and network/gene
statistics}: The three columns are the rank correlation between GeneSurrounder
results ($p^{\text{GS}}$) and network/gene statistics (Degree, Betweenness, and
$p^{\text{DE}}$) across all genes in each dataset. The Degree and Betweenness
are two different network centrality measures. The Degree is the number of
connections a node has and the Betweenness is the fraction of  shortest paths
that passes through the node.  $p^{\text{DE}}$ is the $p$-value obtained from a
standard  differential expression $t$-test.}

\label{tab:bladder_integrationdata}

\end{table}

\setlength{\tabcolsep}{15pt} 

\begin{table}[!hb]
\centering

\begin{tabular*}{\columnwidth}{ l  c  c  c }

  \toprule
  Bladder Cancer Study Pair & $p^{\text{GS}}$ Cor. & $p^{\text{DE}}$ Cor. & $p^{\text{LEAN}}$ Cor. \\
  \midrule
  GSE13507 - GSE31684 & $+ 0.033$  & $+ 0.008$ & $- 0.086$ \\ 
  GSE13507 - GSE32894 & $+ 0.317$ & $- 0.045$ & $- 0.020$ \\
  GSE31684 - GSE32894 & $+ 0.119$ & $+ 0.010$ & $- 0.005$ \\
  \bottomrule
\end{tabular*}

\caption{\textbf{Cross study concordance of GeneSurrounder results compared to
differential expression analysis and LEAN}: The columns $p^{\text{GS}}$ Cor.,
$p^{\text{DE}}$ Cor., and $p^{\text{LEAN}}$ Cor. are the Spearman rank
correlations respectively  between the results obtained from  GeneSurrounder,
differential expression analysis, and LEAN for each study pair. }

\label{tab:bladder_xstudy}

\end{table}









\end{document}